# Simulation of NMR Fermi contact shifts for Lithium battery materials: the need of an efficient hybrid functional approach

Yuesheng Zhang,[1] Aurore Castets,[2,3] Dany Carlier,[2,3] Michel Ménétrier,[2,3] and Florent Boucher[1*]

[1]Institut des Matériaux Jean Rouxel (IMN), Université de Nantes, CNRS,
2 rue de la Houssinière, BP 32229, 44322 Nantes Cedex (France)
[2]CNRS, Université de Bordeaux, ICMCB,
87 avenue du Dr. A. Schweitzer, 33608 Pessac Cedex (France)
[3]CNRS, IPB-ENSCBP, ICMCB,
87 avenue du Dr. A. Schweitzer, 33608 Pessac Cedex (France)

In the context of the development of NMR Fermi contact shift calculations for assisting structural characterization of battery materials, we propose an accurate, efficient, and robust approach based on the use of an all electron method. The full-potential linearized augmented plane wave method, as implemented in the WIEN2k code, is coupled with the use of hybrid functionals for the evaluation of hyperfine field quantities. The WIEN2k code is able to fully relax relativistic core states and uses an autoadaptive basis set that is highly accurate for the determination of the hyperfine field. Furthermore, the way hybrid functional approaches are implemented offers the possibility to use them at no additional computational cost. In this paper, NMR Fermi contact shifts for lithium are studied in different classes of paramagnetic materials that present an interest in the field of Li-ion batteries: olivine Li$M$PO$_4$ ($M$ = Mn, Fe, Co, Ni), anti-NASICON type Li$_3M_2$(PO$_4$)$_3$ ($M$ = Fe, V), and antifluorite-type Li$_6$CoO$_4$. Making use of the possibility to apply partial hybrid functionals either only on the magnetic atom or also on the anionic species, we evidence the role played by oxygen atoms on polarisation mechanisms. Our method is quite general for an application on various types of materials.

PACS : 76.60.-k, 71.15.Mb

---

[*]Electronic addresses: Florent.Boucher@cnrs-imn.fr

I. INTRODUCTION

In the context of finding new and efficient electrochemical energy storage solutions, important efforts have been devoted to the study of transition metal (TM) compounds as materials for positive electrodes in Li ion batteries.[1] In this highly competitive and strongly applied research domain, there is still an important need in having a very good knowledge of the bonding properties of the active material together with information on its crystal structure and/or local structure organization. On the latter point, solid state magic angle spinning nuclear magnetic resonance (MAS-NMR) is a very powerful and accurate technique that is now widely used for studying electrode materials. Relevant examples can be found in Ref. 2-8 where $^7$Li or $^6$Li MAS-NMR is used to study Li environment in battery materials.

Due to the hyperfine interactions between the nuclear magnetic moment of Li and the electronic spin density induced by paramagnetic TM ions, the NMR shift of Li nucleus in Li-TM compounds is often largely different from that in diamagnetic environments. Hyperfine interactions consist of through space dipolar interactions and a through bond Fermi contact one. For Li-TM oxide compounds considered here, the Li MAS-NMR shift is mainly determined by the isotropic Fermi contact interaction which is due to the presence of electron spin density on the Li nucleus, as transferred from the orbitals of neighboring paramagnetic species. When different NMR signals are observed in concordance with the existence of more than one Li environment, it becomes crucial to correctly assign those NMR peaks. When paramagnetic compounds are submitted to an external magnetic field, a macroscopic magnetization develops due to the difference in the population of the two spin channels: the majority one (up, aligned with the applied field) and the minority one (down). Two basic mechanisms have been proposed to account for the transfer of electron spin density inducing the Fermi contact shift in lithium transition metal compounds: spin delocalization and spin polarization.[9] The delocalization mechanism corresponds to transfer of unpaired electron spin density from a TM $d$ orbital to the Li nucleus with the same polarity (up) by orbital overlaps, either directly or via oxygen. This leads to a positive NMR shift on the Li nucleus. The spin polarization mechanism corresponds to the polarization of a fully occupied (either $d$ cationic or $p$ anionic) orbital induced by unpaired electrons in the $d$ orbital of the TM ion. Due to the presence of spin polarized partially occupied d orbitals, mixing interactions between the anionic $p$ and cationic $d$ levels are different on the majority (up) and minority (down) spin channels. This induces a negative spin polarization for the occupied orbitals which can be transferred to Li again either directly or via oxygen depending on the geometry. Consequently, a negative NMR shift on Li ion is observed when such polarization occurs. This mechanism will be discussed in detail in a forthcoming paper. The occurrence and the relative importance of these delocalization and polarization mechanisms depend on the spin state of transition metal ions and the local environment of the Li ion. An analysis based on these two mechanisms can be used to qualitatively assign the lithium NMR signals in simple cases, but it remains difficult for compounds with complicated structure or when only small differences between different signals exist. Hence, theoretical calculations are very helpful for this purpose and some of us initiated such developments.[8-10] Using the VASP code, a qualitative approach based on plane wave/pseudo-potential calculation was thus introduced to investigate the transfer mechanisms. Furthermore, the assignment of the NMR peaks in a given compound was correlated to the integrated spin density found in a sphere around Li. However, as shown below this method suffers from some limitations and, in some cases, correct assignment of NMR shifts requires the calculation of the Fermi contact shift.

If the paramagnetic behavior of the compound obeys ideal Curie law (spin only), the Fermi contact shift of the probed nucleus (lithium for instance) can be expressed as:[11]

$$\delta^i_{con} = \frac{A^i}{\hbar \gamma^i} \frac{g_e \mu_B S(S+1)}{3kT} = \frac{A^i}{\hbar \gamma^i} \frac{\mu^2_{eff-theory}}{g_e \mu_B 3kT} = \frac{A^i}{\hbar \gamma^i} \frac{\chi_{M-Curie}}{N_A \mu_0 g_e \mu_B} \quad (1)$$

where $S$ is the total spin quantum number of the paramagnetic ions, $g_e$ is the free electron g factor, $\mu_B$ is the Bohr magneton, $\mu_0$ is the magnetic permeability in vacuum, $\gamma^i$ is the gyromagnetic ratio of the probed nucleus, $\hbar$ is the reduced Plank constant, $k$ is Boltzman constant, $\chi_{M-Curie}$ is the Curie-like theoretical molar magnetic susceptibility, $\mu_{eff-theory}$ is the theoretical magnetic moment, $N_A$ is Avogadro's number, and $T$ is the temperature. $A^i$ is the isotropic hyperfine coupling constant, which is related to the portion of the global magnetization that is present at the nucleus site:

$$A^i = \hbar \gamma^i \frac{\mu_0}{3S} g_e \mu_B \rho^i_{HFF}(0) \quad (2)$$

where $\rho^i_{HFF}(0)$ is the electron spin density averaged inside the Thomson sphere and has been used by many authors for the calculation of hyperfine field:[12-14]

$$HFF^i = \frac{8\pi}{3} \mu_B \rho^i_{HFF}(0) \quad (3)$$

Whereas Eq. (1) has been used recently by Mali et al. for the calculation of the Fermi contact shift in orthosilicate positive electrode materials,[15,16] the experimental magnetic susceptibility often deviates from the theoretical Curie-type susceptibility due to possible orbital contribution and/or residual magnetic correlations. In such situation, the Fermi contact shift can be calculated using either the experimental effective magnetic moment or the magnetic susceptibility as:

(*i*) if a Curie-Weiss behavior is obeyed

$$\delta^i_{con} = \frac{A^i}{\hbar \gamma^j} \frac{\mu^2_{eff-\exp}}{g_e \mu_B 3k(T-\Theta)} \quad (4a)$$

(*ii*) in a general case

$$\delta^i_{con} = \frac{A^i}{\hbar \gamma^j} \frac{\chi_{M-\exp}}{N_A \mu_0 g_e \mu_B} \quad (4b)$$

where $\chi_{M-exp}$ is the experimental molar magnetic susceptibility, $\mu_{eff-\exp}$ is the experimental effective magnetic moment, $\Theta$ is the Weiss constant.

A recent paper of Kim et al.[17] clearly recommends the use of the experimentally determined Curie-Weiss parameters to calculate the Fermi contact shift, which is also supported by our previous calculations.[18] Moreover, we showed that for cases where the magnetic behavior is not Curie-Weiss type at the temperature of the NMR measurements, one should preferably use the experimental susceptibility at this temperature to calculate the NMR shift.[19] Therefore, in the present paper, we will only deal with Fermi contact shift values calculated with Eqs. (4a) or (4b) depending of the applicability of the Curie-Weiss law or the knowledge of the experimental susceptibility. Furthermore, the temperature used in these equations will be that of the sample during the NMR measurement with which we wish to make a comparison. Based on the details of the NMR measurements, this temperature was estimated as stated in each case.

Assuming that the magnetic susceptibility is known, the other key quantity that enters Eq. (4b) for the calculation of Fermi contact is the hyperfine coupling constants $A^i$ for which an accurate knowledge of $\rho_{HFF}^i(0)$ is needed. Many calculations of Fermi contact coupling constants of atoms or molecules have been performed with post Hartree-Fock theory using molecular approaches.[20-24] Those results showed that it is a great challenge to obtain a good agreement between calculations and experiments, the accuracy of these calculations being very much dependent of the basis set completeness and of the correlation level. Pretty good agreement between experimental and theoretical values can be obtained using high level post Hartree-Fock methods coupled with extended basis sets.[22-24] Unfortunately, such calculations are limited to small molecular or cluster systems due to the large cost of such computations. In recent years, Density Functional Theory (DFT) has been considered as a good alternative to post Hartree-Fock methods for the calculation of Fermi contact coupling constants as it requires a much lower computational cost. Many examples of applications can be found on molecular systems with lots of tests and developments using various exchange/correlation functionals or for improving the basis set.[25-31] Concerning the study of crystalline compounds, people in the context of Mössbauer or ESR spectroscopies were traditionally more interested in the hyperfine field (HFF) which is indeed a quantity directly correlated to the spin density around the nucleus (Eq. (3)).

While all-electrons methods are still the reference ones when periodic boundary conditions need to be applied,[12,32-35] pseudo-potential approaches have recently demonstrated their efficiency for the calculation of $\rho^i_{HFF}(0)$.[36] This is related to the development of the PAW implementation[37] and, in addition, the possibility to take into account the core states relaxation.[38-40]

In DFT calculations, the quantum correlation effect (exchange and correlation interaction) between electrons is taken into account explicitly (though approximately), which is important for the calculation of hyperfine coupling constants. However, the exact form of exchange-correlation potential in DFT is unknown, so approximations have to be made for DFT to become a practical tool. The local density approximation (LDA) and generalized gradient approximation (GGA) to the exchange-correlation potential have been widely used in DFT calculations. However, there is one basic deficiency in LDA or GGA: the self interaction error (SIE) of each electron in the Hartree term is not completely canceled by the exchange-correlation term. There is no such error in Hartree-Fock theory because this self interaction can be exactly canceled by the Fock exchange interaction term ("exact" exchange potential). The SIE has a large effect on localized transition-metal $d$ or rare earth $f$ orbitals, hence LDA or GGA is not appropriate to describe these orbitals. To overcome this deficiency, many methods have been proposed. One is the "GGA+$U$". In this approach, Hubbard type interaction is added to localized $d$ or $f$ electrons.[41] The other method that is now widely used is the hybrid functional method: a portion of LDA or GGA exchange potential is replaced by Fock-type exact exchange potential to remove the self interaction error.[42] Thus, a better description for localized $d$ or $f$ orbitals can be expected from it.

In order to obtain correct hyperfine coupling constants $A^i$, we need a method that can accurately describe the valence states but also the semi-core and core states very close to the nucleus. Mali et al. have calculated hyperfine coupling constants of $Li_2MnSiO_4$ polymorphs with the Quantum Espresso package. They used standard GGA for the exchange correlation potential, a plane wave basis set coupled with norm-conserving pseudopotentials, and the PAW development implemented in the "GIPAW package" for recovering all-electrons properties but apparently no core states relaxation. NMR Fermi contact shifts were calculated by assuming an ideal Curie behavior (Eq. (1)) at room temperature.[15,16] Kim et al. have adopted a more accurate and robust approach for studying several paramagnetic Fe(III) phosphates.[17] They have used an all-electron linear combination of atomic orbitals (LCAO) in the Crystal06 code together with hybrid functional method to calculate the spin density at the nucleus. NMR Fermi contact shifts were then obtained considering experimental Curie-Weiss type behaviors (Eq. (4a)). Such an approach is however very computer-demanding as the Fermi contact shift calculation requires large basis sets for having a good accuracy.

In the present paper, we selected the full-potential linearized augmented plane wave (FP-LAPW) method implemented into the WIEN2k code to perform the calculations,[43] as we recently did for several tavorite-like $LiMPO_4.OH$ and $MPO_4.H_2O$ phases.[18,19] The full-potential linearized augmented plane wave (FP-LAPW) method is usually recognized to be the most precise one for electronic structure calculation of periodic compounds. Contrary to an LCAO method, the Full Potential WIEN2k/LAPW method uses an adaptive basis set whose completeness can be easily evaluated. Furthermore, the possibility to use a fully relativistic basis set has demonstrated its interest in the determination of nuclear properties like HFF[13] (or $\rho_{HFF}(0)$, Eq. (3)) which is obtained by averaging spin density inside the nuclear Thomson sphere.[12] Beside the standard GGA method, we also evaluated the performance of GGA+$U$ and hybrid functional methods within this code. In WIEN2k, the unit cell is divided into (1) non-overlapping atomic spheres centered at the atomic sites and (2) an interstitial region. The "$U$" term in GGA+$U$ is applied to localized $d$ or $f$ orbitals, only inside the corresponding atomic spheres. This is the general way to perform the GGA+$U$ calculation. For hybrid functional calculations, the situation is more complex. The partial exchange potential is usually applied in the whole unit cell, but in WIEN2k, the implementation of hybrid functional methods uses a similar strategy as that of GGA+$U$. Therefore, partial exact exchange potential is also applied inside the selected atomic spheres only, which makes its calculation cost comparable with that of GGA or GGA+$U$, and its result to be similar with GGA+$U$ to some extent.

In this article, we have selected several Li transition-metal compounds to calculate the Fermi contact shift of the Li nucleus. First, we selected the Li$M$PO$_4$ ($M$=Mn, Fe, Co, Ni) phosphates with olivine structure. There is only one kind of Li site in Li$M$PO$_4$, and the experimental Fermi contact shifts for Li are positive (Mn) and negative (Fe, Co and Ni), respectively.[3] Second is the antifluorite-type Li$_6$CoO$_4$. The Fermi contact shifts of the two Li sites have been measured and qualitatively analyzed by some of us before.[8]

Here we will present quantitative results and compare them with the previous ones. Finally, the monoclinic $Li_3Fe_2(PO_4)_3$ and $Li_3V_2(PO_4)_3$ phases are discussed. The Fermi contact shifts of the three Li ions in $Li_3V_2(PO_4)_3$ have been measured by Cahill et al.,[6] and that of $Li_3Fe_2(PO_4)_3$ have been measured by Davis et al.,[7] and simultaneously by some of us.[10] The Fermi contact shifts and mechanisms in these two compounds have been qualitatively modeled by some of us.[10] This will allow us to point out the limitations of the earlier qualitative method.

## II. COMPUTATIONAL DETAILS

Before calculating $\rho_{HFF}(0)$ (or HFF) with the WIEN2k code, the crystal structure of every compound has been fully relaxed with the Vienna ab initio Simulation Package (VASP),[44-46] using the experimental crystal structure as a starting point. Relaxations have been carried out using the PBE-type GGA[47] for the exchange correlation potential and all the $\rho_{HFF}(0)$ calculations obtained afterwards with different functionals (GGA, GGA+U, Hybrid) are performed on these optimized structures. By this way, we can directly compare the effect of different exchange-correlation potentials on the spin transfer mechanism keeping the structural features unchanged. Bonding and electronic structure analysis can be done afterwards and will be detailed on a forthcoming paper.[48] To test the validity of our calculations, HFF were also calculated on structures relaxed within the GGA+U formalism. In most cases, the conclusions obtained with these two different sets of structures are similar, hence only the results obtained by the former methodology (GGA relaxation) are included in this paper. The relaxation has been performed using the standard PAW PBE pseudopotentials proposed with the VASP package, a plane wave energy cut-off of 600eV and a k-mesh dense enough to reach the convergence.

For the $\rho_{HFF}(0)$ calculations with WIEN2k, the spherical radii of atoms have been chosen in such a way that those spheres are nearly touching in the unit cell. The number of k-points in the Brillouin zone and the energy cutoff have been chosen to ensure the convergence of total energy, charge and spin density $\rho_{HFF}(0)$. Thus, the parameters used in the calculation are (1) Li$M$PO$_4$ ($M$=Mn, Fe, Co, Ni): $R_{min}K_{max} \approx 7.0$; atomic spheres radii of Li, P and $M$(Mn, Fe, Co, Ni) are 1.87, 1.61 and 2.0 a.u., respectively; atomic sphere radii of O in $M$=Mn, Fe, Co, Ni are 1.3, 1.26, 1.3, 1.27 a.u., respectively; 47 irreducible k-points with a (5×9×11) k-point mesh ; (2) Li$_6$CoO$_4$: $R_{min}K_{max}$ =7.6, 90 irreducible k-points with a (9×9×12) k-point mesh, atomic spheres of Li, Co and O are 1.87, 1.99 and 1.7 a.u., respectively; (3) Li$_3M_2$(PO$_4$)$_3$ ($M$=Fe, V): $R_{min}K_{max}$ = 6.9; 24 irreducible k-points with a (5×3×5) k-point mesh; the atomic spheres of Li, V, P and O in V case are 1.80, 1.87, 1.44 and 1.44 a.u., respectively; while the atomic spheres of Li, Fe, P and O in Fe case are 1.81, 1.89, 1.42 and 1.42 a.u., respectively.

As mentioned previously, the purpose of this study is to quantitatively evaluate NMR Fermi contact shifts in order to discuss on the performance of various functional for the accurate description of the $\rho_{HFF}(0)$. Therefore, we chose a set of compounds presenting different spin transfer mechanisms. Beside standard PBE/GGA approach, GGA+$U$ and hybrid functional methods were tested. The Hubbard $U$ correction was applied on the TM-3$d$ orbitals using the formalism introduced by Anisimov et al.[49] The adequate value for the "$U$" term being however unknown for most TM; we take it as a variable in the calculation. Only an effective value $U_{eff}=U-J$ is used in the present GGA+$U$ method, where $U$ is the on site electron-electron repulsion and $J$ is the exchange interaction. Concerning hybrid functionals, there are many kinds of them in the literature, such as PBE0, B3LYP, B3PW91, HSE, etc., and every method contains a fixed weight of exact exchange potential. However, the suitable ratio of exact exchange potential to be applied depends on the system. Thus, it is not easy to get good results for all compounds with a fixed weight of exact exchange potential, especially for the calculation of Fermi contact shifts, which are subtly dependent on the electronic states of neighboring ions. Here, we therefore adopt the PBE-Fock-α method for the hybrid functional calculations. In this method, the weight of exact exchange potential can be controlled by the fractional variable $α$.

The exchange-correlation potential $E_{xc}^{PBE-Fock-α}$ is calculated as:

$$E_{xc}^{PBE-Fock-α} = E_{xc}^{PBE}[\rho] + α \left( E_x^{HF}[\psi_{corr}] - E_x^{PBE}[\rho_{corr}] \right) \qquad (5)$$

where $E_{xc}^{PBE}[\rho]$ is the PBE exchange-correlation potential of the system, $E_x^{HF}[\psi_{corr}]$ and $E_x^{PBE}[\rho_{corr}]$ are exact and PBE-type exchange potential of the corresponding orbitals.

In the WIEN2k code, the exact-exchange/hybrid methods are implemented only inside the atomic spheres and traditionally for the localized electrons. However, as we have mentioned above, the oxygen ion may play an important role in the spin transfer from the TM ions to the Li nucleus, via both spin delocalization and spin polarization mechanisms. To take this into account, we have also applied the Fock correction to O-2$p$ orbitals in addition to TM-3$d$ orbitals. Thus, there are two types of Hybrid

Functional calculations in this paper: in the first case, the partial GGA exchange potential of the sole transition-metal $3d$ orbitals inside the corresponding atomic spheres is replaced by the corresponding exact potential (Hybrid Functional I: HyF1) and in the other case the partial GGA exchange potential of transition-metal $3d$ and O-$2p$ orbitals in the corresponding atomic spheres are both replaced by the corresponding exact potential (Hybrid Functional II: HyF2). This distinction is important in the calculation as we will see in the following. Furthermore, partial GGA exchange potential of other atoms in the unit cell can also be replaced by exact exchange potential besides transition-metal $3d$ and O-$2p$ orbitals, but the results show that there is no improvement in this calculation when compared to HyF2 kind of calculation (as illustrated for selected examples in the supplementary information, Table S-I).

### III. RESULTS AND DISCUSSSION

#### A. Study of Li$M$PO$_4$ phases ($M$=Mn, Fe, Co, Ni)

Olivine Li$M$PO$_4$ phases crystallize in the *Pnma* space group, and consist of distorted LiO$_6$, $M$O$_6$ and PO$_4$ units. The phosphorous ions occupy tetrahedral sites, while the lithium and transition-metal ions occupy octahedral sites. There is only one kind of lithium ion site ($4a$ site) in Li$M$PO$_4$ compounds. Every LiO$_6$ octahedron shares edges with two $M$O$_6$ octahedra and corners with four other $M$O$_6$ polyhedra (Fig. 1). The experimental crystal parameters are: (LiMnPO$_4$: a=10.431 Å, b=6.0947 Å, c=4.7366 Å),[50] (LiFePO$_4$: a=10.338 Å, b=6.011 Å, c=4.695 Å),[51] (LiCoPO$_4$: a=10.2001 Å, b=5.9199 Å, c=4.690 Å),[52] and (LiNiPO$_4$: a=10.0275 Å, b=5.8537 Å,

c=4.6763 Å).[50] The transition metal ions in these compounds are divalent: High Spin(HS)-$Mn^{2+}$ ($3d^5$: $t_{2g}^3 e_g^2$, S=5/2), HS-$Fe^{2+}$ ($3d^6$: $t_{2g}^4 e_g^2$, S=2), HS-$Co^{2+}$ ($3d^7$: $t_{2g}^5 e_g^2$, S=3/2) and $Ni^{2+}$ ($3d^8$: $t_{2g}^6 e_g^2$, S=1). For convenience, here we still use the term "$t_{2g}$" and "$e_g$" to denote electrons in different d orbitals of M ions though it is not strictly correct because of the distortion of $MO_6$ octahedra. Note that we also adopt the usual denomination "$e_g$" for the strictly speaking $e_g^*$ antibonding hybrid orbital with oxygen. The experimental room temperature NMR Fermi contact shifts of $^7Li$ at 310 K, effective magnetic moments and Weiss constant for $LiMnPO_4$, $LiFePO_4$, $LiCoPO_4$ and $LiNiPO_4$ are (68 ppm, 5.4 $\mu_B$, -58 K), (-8 ppm, 6.8 $\mu_B$, -161 K), (-86 ppm, 5.1 $\mu_B$, -70 K), and (-49 ppm, 3.1 $\mu_B$, -60 K), respectively.[3] We used Eq. (4a) to calculate the Fermi contact shift of Li, listed in Table I and plotted in Fig. 2, with T=310 K.

For the GGA case, the calculated values are 131 *ppm* (Mn), 118 *ppm* (Fe), 73 *ppm* (Co) and -63 *ppm* (Ni), respectively. The calculated value for Li in $LiNiPO_4$ is negative and only a little smaller (i.e. larger in absolute value) than the experimental one. On the contrary, the calculated values of $LiMnPO_4$, $LiFePO_4$ and $LiCoPO_4$ are much larger than the corresponding experimental ones. In fact, the negative chemical shifts of Li in $LiFePO_4$ and $LiCoPO_4$ are even predicted to be positive by the GGA calculation. The selected results of "GGA+U" method are also listed in Table I and plotted in Fig. 2, and are better than those from GGA calculation. For example, the calculated Fermi contact shift of Li in $LiNiPO_4$ are -53 *ppm* (with $U_{eff}$=0.1 Ry) and -40 *ppm* (with $U_{eff}$=0.3 Ry), respectively, which are close to the experimental values. This implies that the U term on Ni-3d decreases the spin polarization effect from $e_g$ orbitals to the Li nucleus. The positive Fermi contact shift of LiMnPO$_4$ is decreased with the increase of $U_{eff}$ and consequently the agreement between calculation and experiment is improved. However, for $U_{eff}$=0.5 Ry, the calculated value is about 93*ppm*, which is still larger than the experimental value of 68*ppm*. Better agreement with experiment can be obtained using even larger $U_{eff}$(0.7Ry~1.0Ry), but this is over the values commonly accepted for transitional metal ions. For LiFePO$_4$ and LiCoPO$_4$, the calculated Fermi contact shifts are also decreased with the increase of $U_{eff}$, but they are still positive, even when very large $U_{eff}$ is used in the calculation. An important result is therefore that we cannot get a negative Fermi contact shift for Li in these two compounds with GGA+$U$ calculations.

No improvement over the GGA+$U$ method can be obtained when partial exact exchange potential is added to the transition-metal ions only (HyF1). From Table I and Fig. 2, we find that the result of HyF1 calculation is similar to that of GGA+$U$: the best agreement with experiment can be obtained in LiNiPO$_4$, then LiMnPO$_4$, while the (experimentally negative) Fermi contact shifts in LiFePO$_4$ and LiCoPO$_4$ are still predicted to be positive. In contrast, obvious improvement can be obtained with the HyF2 method. Now the positive Fermi contact shift of Li ion in LiMnPO$_4$ decreases largely with the increase of exact exchange potential, and good agreement with experiment can be obtained when ~35% of exact exchange potential (mixing parameter α=0.35) is added. The negative value of LiNiPO$_4$ increases slowly (smaller absolute value) when a large portion of exact exchange potential is added, and a pretty good agreement with experiment is also obtained with a mixing parameter α≈0.35. The most obvious improvement is observed in LiNiPO$_4$ and LiCoPO$_4$. The experimental negative

Fermi contact shifts are now predicted by the HyF2 calculations with the appropriate portion of exact exchange potential: it is about -2 *ppm* with α≈0.5 for LiFePO$_4$ and -21 *ppm* with α≈0.35 for LiCoPO$_4$. However, these negative values do not decrease monotonously with an increase of exact exchange potential, as can be seen from the results of α=0.35 and 0.5. We propose that the difference between HyF2 and HyF1 methods for the Fermi contact calculation can be understood as follows. Though the SIE deficiency in GGA is generally strong for localized *d* or *f* orbitals, it should also have some influence on other orbitals, such as O-2$p$ ones. These influences may be tiny on the calculation of common quantities, such as total energy and magnetic moment, and can be omitted in the corresponding calculation. However, for a quantity sensitive to neighboring environment, such as Fermi contact shift, these influences can not be omitted. Thus, to improve the results of the Fermi contact shift calculations, the addition of partial exact exchange potential is also needed for the O-2$p$ orbitals. From the calculated Fermi contact shifts of Olivine Li*M*PO$_4$ compounds with HyF2 method, we find that the best agreement with experiment can be obtained when mixing parameter α is about 0.35 except for LiFePO$_4$. The calculated Fermi contact shift of Li in LiFePO$_4$ is about 9 ppm when α=0.35 (-8 ppm experimentally). Though they have opposite signs, the difference between the calculated and experimental values is not large, and we can conclude that a mixing parameter α=0.35 is globally appropriate to calculate the Fermi contact shift of Li ions in Olivine Li*M*PO$_4$ compounds.

The spin polarized density of states (DOS) and magnetic moment on transition metal ions for Li*M*PO$_4$ compounds are given in supplementary information (Table S-II and Fig. S-1), showing that the expected transition metal spin states mentioned above are obtained whatever the calculation. It is interesting to note that the distortion of the TM octahedra is strong enough to induce a separation in energy within the $t_{2g}$ orbitals. This can be seen from the projected DOS on Co-3$d$ orbitals in LiCoPO$_4$ in the supplementary information Fig. S-2 and S-3, where two spin-down electrons mainly occupy spin-down $d_{xz}$ and $d_{yz}$ orbitals and leave the spin-down $d_{xy}$ orbital empty. This is an important fact for analyzing the electron spin density transfer mechanisms, as will be illustrated in subsequent reports.

There is obvious difference between the theoretical (spin-only) and experimental effective magnetic moments of Li$M$PO$_4$, especially for LiFePO$_4$ and LiCoPO$_4$.[3] In order to consider whether the orbital magnetic moment that may contribute to this difference can affect the Fermi contact shift on Li ions, we also performed the GGA or GGA+$U$ calculation including spin-orbital coupling (SO). The result can be found in the supplementary information Table S-III: the difference of Fermi contact shift of Li obtained by calculation with and without spin-orbital coupling is negligible for Li$M$PO$_4$. Similar conclusion can also be found in the Li$_6$CoO$_4$ case discussed in the following section.

### B. Study of the Li$_6$CoO$_4$ phase

The space group of Li$_6$CoO$_4$ is tetragonal P4$_2$/nmc. The crystal structure contains CoO$_4$ tetrahedra and two different types of Li sites: Li(1) and Li(2), where the Li(1)O$_4$ and Li(2)O$_4$ tetrahedra present different bonding topologies with the surrounding atoms (Fig. 3). The Li(1)O$_4$ tetrahedron shares one edge and two corners with CoO$_4$ tetrahedra, while the Li(2)O$_4$ tetrahedron shares its four corners with four CoO$_4$ tetrahedra. The lattice parameters of Li$_6$CoO$_4$ were fully relaxed with VASP (GGA) using the experimental ones as a starting point ($a$=6.536 Å and $c$=4.654 Å).[53] Co$^{2+}$ ions in Li$_6$CoO$_4$ are in $e^4t_2^3$ configuration ($S$=3/2). Some of us have recorded $^7$Li NMR spectra showing two isotropic signals corresponding to the two Li sites: one is strongly positive (885 ppm) and the other is strongly negative (-232 ppm). In this study, the electronic structure of Li$_6$CoO$_4$ was also calculated using the VASP code to help in the interpretation of NMR experimental results.[8] However, because the spin density near the Li nucleus or the HFF is not currently accessible in VASP, the net spin inside spheres around each Li ion (calculated by integrating the spin density inside those spheres) was used for interpreting the Fermi contact shift. With this method, the positive and negative shifts were unambiguously assigned to Li(2) and Li(1), respectively, but the calculated Li(2)/Li(1) net spin ratio was largely different from the experimental Fermi contact shifts ratio. Indeed, the experimental Fermi contact shift ratio Li(2)/Li(1) is close to 3.8 (in absolute value), whereas the spin densities calculated with GGA in a 0.6 Å radius sphere yield a much larger ratio of 14.8. With GGA+$U$ method, the Li(2)/Li(1) absolute spin density ratio was lowered (12.2), but it is still much larger than that of the NMR shifts.

  The Fermi contact shifts of Li$_6$CoO$_4$ is calculated using Eq. (4a) and considering the experimental Weiss constant of -8 K and the effective magnetic moment of 4.1 $\mu_B$.[54]. 320 K is taken for the sample temperature in the 30 kHz spinning NMR measurement, as determined by standardization.[19] The results are given in Table I and shown in Fig. 4. With GGA calculation, the calculated Fermi contact shifts on Li(1) and Li(2) are about -176 *ppm* and 1441 *ppm*, respectively. Both values have the correct sign, *i.e.* similar to the corresponding experimental ones (-232 *ppm, and* 885 *ppm*), but the calculated shift for Li(2) is too large and for Li(1) it is not enough negative. Consequently, while the calculated ratio between those two values has been improved compared to previous calculations,[8] it is still too large (in absolute value) compared to the experimental one (8.2 against 3.8 respectively). Applying "GGA+$U$" method to Co-3$d$ electrons, the positive Fermi contact shift on Li(2) is decreased, and a better agreement with the experiment is obtained. However, the negative Fermi contact shift on Li(1) is also decreased (in absolute value), so that the difference between calculation and experiment is even larger than in the GGA case. Therefore, the ratio between the calculated NMR shifts of Li(2) and Li(1) is still far from the experimental one whatever the $U_{eff}$ term introduced in the calculation: 8.1, 9.9, and 7.2 (in absolute value) with $U_{eff}$=0.1, 0.3 and 0.5 Ry, respectively. Just like for the Li$M$PO$_4$ compounds, adding partial exact exchange potential for the Co-3$d$ orbitals (HyF1) has an effect similar to the GGA+$U$ method: both the positive and negative NMR shifts are weakened compared to the pure GGA functional. Consequently, only the Fermi contact shift calculation on Li(2) can be improved, while a worse disagreement with experiment is obtained for the negative shift of Li(1). On the other hand, if we add partial exact exchange potential to both Co-3$d$ and O-2$p$ orbitals (HyF2), a good agreement with experiment can be obtained: the positive value on Li(2) is decreased and the negative value on Li(1) is enlarged, leading to a much improved global agreement. The best agreement with experiment for both Li(1) and Li(2) can be obtained when about 15% of the GGA exchange potential is replaced by the corresponding exact potential: 941(Li(2)) and -265(Li(1)) with a ratio close to 3.5 (in absolute value). To check the validity of each of these calculations, we also plotted the spin-polarized DOS and computed the magnetic moment on Co (in supplementary information Table S-II and Fig. S-4), which shows that $Co^{2+}$ ($e^4 t_2^3$) is obtained in each case. To summarize this part and as already observed for the olivine phases, the anionic p states play an important role in the polarization mechanism and therefore have to be corrected from the SIE using a hybrid scheme.

C. Study of the $Li_3M_2(PO_4)_3$ (*M*=V, Fe) phases

The two $Li_3Fe_2(PO_4)_3$ and $Li_3V_2(PO_4)_3$ phases are isostructural with a monoclinic cell ($P2_1/n$ space group). The lithium ions are distributed in three distinct sites: Li(1), Li(2) and Li(3) (Fig. 5). Here we use the nomenclature adopted by Patoux *et al.*[55] Li (1) is in a tetrahedral site sharing edges with two $MO_6$ octahedra; Li(2) is in a distorted trigonal bipyramid site sharing respectively an edge and a face with two $MO_6$ octahedra; finally, Li(3) is also in a distorted trigonal bipyramid site sharing respectively a face and two corners with three $MO_6$ octahedra. There are two non symmetry related positions of Fe or V in $Li_3M_2(PO_4)_3$. They occupy distorted octahedral sites with the following configurations: HS $Fe^{3+}$($t_{2g}^3 e_g^2$) and $V^{3+}$($t_{2g}^2 e_g^0$). The $^7Li$ MAS NMR spectra of $Li_3Fe_2(PO_4)_3$ have been measured by some of us with the observation of three distinct signatures at 189 ppm, 89 ppm, and 39 ppm, respectively.[10] Similar results have been simultaneously obtained by Davis et al.[7] The $^7$Li NMR spectrum of $Li_3V_2(PO_4)_3$ have been recorded by Cahill et al. and exhibits three isotropic signals at 103 ppm, 52 ppm and 17 ppm, respectively.[6] The unambiguous assignment of NMR signals to the three distinct Li sites is however not obvious due to the complicated crystalline structures and the small differences observed between those NMR signals. For instance, in the case of the $Li_3Fe_2(PO_4)_3$ phase, Davis et al. proposed the following ranking for the three Li NMR peaks: $\delta$ (Li3) > $\delta$ (Li1) > $\delta$ (Li2), based on the structural analysis,[7] while Kim et al. gave a different assignment using calculations based on the Crystal06 code: $\delta$ (Li1) > $\delta$ (Li2) > $\delta$ (Li3).[17] For the $Li_3V_2(PO_4)_3$ phase, Cahill et al. proposed a ranking similar to the one suggested by Davis et al. on $Li_3Fe_2(PO_4)_3$ *i.e.* $\delta$ (Li3) > $\delta$ (Li1) > $\delta$ (Li2).[6] In our previous paper where the VASP code is used for interpreting the NMR Fermi contact shift, different attributions were proposed for these two compounds.[10] Effectively, as in the case of $Li_6CoO_4$, the HFF or spin density near the Li nucleus was not directly accessible. Consequently, the net spin (or integrated spin amount) inside spheres around each Li ion was used to represent the Fermi contact shift. For $Li_3Fe_2(PO_4)_3$, these calculations led to the following assignments: $\delta$ (Li1) > $\delta$ (Li3) > $\delta$(Li2), whatever the method (GGA or GGA+$U$) or the radius of the integration sphere, with rather close expected shifts for Li(3) and Li(2). For $Li_3V_2(PO_4)_3$, the relative net spins calculated around each Li ion depend strongly on the method (GGA or GGA+$U$) used, and within a given method depend very slightly on the radius size. The GGA method leads to the following assignment $\delta$ (Li3) > $\delta$ (Li2) > $\delta$ (Li1), whereas GGA+$U$ leads to the $\delta$ (Li1) > $\delta$(Li3) > $\delta$ (Li2) or $\delta$ (Li3) > $\delta$ (Li1) > $\delta$ (Li2) depending on the radius size and the U value. These results are therefore different from those of Davis et al. or Cahill et al. To further study this question we recalculated the Fermi contact shift of these three Li ions using the WIEN2k approach discussed in the present paper. For $Li_3Fe_2(PO_4)_3$, we have used the experimental effective magnetic moment (5.89 μB) and the Weiss constant (-55 K)[56] to do the calculation (Eq. (4a)), considering 320 K as the sample temperature. The results are presented in Table I and shown in Fig. 6a. The GGA calculated Fermi contact shifts of Li(1), Li(2) and Li(3) in $Li_3Fe_2(PO_4)_3$ are thus 391 ppm, 222 ppm and 196 ppm ($\delta$ (Li1) > $\delta$ (Li2) > $\delta$(Li3)), respectively. Adding a $U$ term or replacing partial GGA exchange-potential with exact exchange potential on Fe-3$d$ orbitals decreases these values and maintains the ranking among them, but the corresponding results are still larger than the experimental values. When partial exact potential is added to both Fe-3$d$ and O-2$p$ orbitals, the situation gets better as we have encountered for the other systems discussed above. There is a reasonable agreement with experiment when about 20% of GGA exchange potential of Fe-3$d$ and O-2$p$ orbitals is replaced by exact potential. With this adjustment, the three signals 189 ppm, 89 ppm and 39 ppm can be assigned to Li(1), Li(2) and Li(3), respectively. The assignment we now propose is therefore different from that obtained with our previous calculations.[10] This difference can be explained by analyzing the evolution of the averaged spin density inside the different atomic spheres. The values reported in Fig. 7a are those obtained by the present WIEN2k calculations with the GGA and GGA+$U$ with a $U_{eff}$=3 eV (this $U_{eff}$=3 eV or 0.22 Ry was one of the $U_{eff}$ values used in Ref. 10). Crossing points can be observed between the spin density curves of Li(2) and Li(3). At small radius (about R<0.45 Å), the averaged spin density on Li(2) is larger than that on Li(3), while at larger radii (R>0.45 Å), the averaged spin density on Li(2) becomes the smallest one. In these previous calculations the radii selected for the analysis were above the crossing point value (between 0.6 Å, and 0.8 Å); this explains the ranking that was proposed: $\delta(Li1) > \delta(Li3) > \delta(Li2)$. Furthermore, this ranking was supported by a detailed analysis of the local configuration for each Li site based on the possible overlap of the Li 2s orbital with idealized 3d orbitals of $Fe^{3+}$ containing the electron spins. Note, however, that the relative positions of Li(2) and Li(3), both from the geometry analysis and the calculations, were considered rather close, as confirmed by the present results in figure 7a. In the present paper, using the HFF approach (Eq. (4a)) to calculate the Fermi contact shift, we now get an assignment consistent with that by Kim et al.[17] ($\delta(Li1) > \delta(Li2) > \delta(Li3)$) with however a better agreement with experimental results.

Turning to the vanadium phase, the experimental molar magnetic susceptibility of $Li_3V_2(PO_4)_3$ at the temperature of the NMR experiment is about 0.006emu/mol.[57] However, the number of V ions is $2N_A$/mole, so the magnetic susceptibility used in the calculation is half of the experimental value. In GGA calculation, the Fermi contact shifts on Li(1), Li(2) and Li(3) ions are 135 ppm, 40 ppm and 65 ppm, respectively and the ranking of Fermi contact shifts is $\delta(Li1) > \delta(Li3) > \delta(Li2)$ (see Table I and Fig. 6b). Adding partial exact exchange potential or $U$ term (GGA+$U$) to Fe-3$d$ orbitals decreases these values, especially that on Li(3). So the calculated Fermi contact shift on Li(3) becomes smaller than that one on Li(2), and now the ranking is $\delta(Li1) > \delta(Li2) >$ $\delta$(Li3). The best agreement with experimental NMR shifts is again obtained for the HyF2 calculation with 20% GGA exchange potential on V-3$d$ and O-2$p$ orbitals replaced by exact exchange potential. Consequently, the three experimental peaks of 103 ppm, 52 ppm and 17 ppm can be assigned to Li(1), Li(2) and Li(3) ions, respectively. This is in agreement with our earlier local geometry analysis that we were not able to confirm using DFT/VASP calculations.[10] The reason of this difference can be analyzed as above for $Li_3Fe_2(PO_4)_3$ looking at the evolution of the average spin density as a function of the sphere radii (see Fig. 7b). When increasing the sphere radius for the integration, we indeed observe the occurrence of crossing points ($\approx$ 0.5-0.6 Å) where Li(2) and Li(3) average spin densities become larger than the Li(1) one. Furthermore, the choice of the functional also determines the existence of a crossing point for Li(2) and Li(3). This clearly demonstrates the limitations of the qualitative method that was previously used for interpreting the NMR Fermi contact shift. In any cases, the best agreement with experiment is obtained for both $Li_3Fe_2(PO_4)_3$ and $Li_3V_2(PO_4)_3$ when 20% GGA exchange potential is replaced by exact exchange potential.

The spin polarized density of states (DOS) and magnetic moment on transition metal ions for all $Li_3Fe_2(PO_4)_3$ and $Li_3V_2(PO_4)_3$ calculations are given in supplementary information Table S-II and Fig. S-5. In $Li_3V_2(PO_4)_3$, there are two $t_{2g}$ electrons for the three $t_{2g}$ orbitals. Our previous calculation with VASP showed that the three $t_{2g}$ orbitals are all partially occupied in GGA calculation, but only two orbitals are occupied in GGA+$U$ calculation,[10] thus participating differently to the spin transfer mechanisms.

To confirm these results, we also plotted the partial V-3$d$ DOS with GGA+$U$($U_{\text{eff}}$=3eV), (see supplementary information Fig. S-6).

IV. CONCLUDING REMARKS

In order to reproduce the experimental results, we have calculated the $^7$Li ion Fermi contact shift in several Li paramagnetic compounds with first principle calculations based on DFT. By studying different structural types or electronic configurations for the transition metal ions, we came to the conclusion that the calculated Fermi contact shift is very much dependant of the exchange correlation potential used in the calculation. Indeed, GGA generally overestimates the spin delocalization mechanism, and on the contrary, underestimates the polarization effects. By using GGA+$U$ or hybrid functional method, one can reasonably well decrease the overestimation for the first mechanism but the best agreement with experiment can only be obtained when partial exact exchange potential is added to both transition metal 3$d$ and O-2$p$ orbitals. The appropriate weight of exact exchange potential in the total exchange potential depends to some extent on the compound, but our calculations show that the same value can be used in similar compounds to get a reasonable Fermi contact shift (≈35% for Li$M$PO$_4$, ≈15% Li$_6$CoO$_4$ for and ≈20% for Li$_3M_2$(PO$_4$), respectively). These results help us to assign the peaks of Li ions Fermi contact shift to different Li sites quantitatively.

From the present work, we can conclude that the F(L)APW method, as implemented in the WIEN2k code, is particularly well adapted to provide the accurate HFF quantities that are needed for the Fermi contact shift calculations. It offers many advantages compared to the different approaches that have been recently proposed in the literature for carrying out similar calculations. Among those advantages, we can recall the benefit of using an auto-adjustable basis set that is accurate enough, and at no additional computational cost, to provide a very precise spin density at the nucleus. Compared to a pseudo-potential/plane wave basis set for which the core state relaxation is not yet fully supported or even implemented or the usage of an LCAO one for which a large basis set and lot of experience and testing is needed before reaching convergence of the HFF, we evidence that the (L)APW basis is particularly well designed and efficient for calculating nuclear-related properties. Furthermore, the judicious way the calculation of exact potential is implemented in the WIEN2k code allowed us to use hybrid functional approaches on large systems without any significant additional computational cost.

## ACKNOWLEDGMENTS

This work benefited from a grant from Agence Nationale de la Recherche (ANR-09-BLAN-0186-01). The Mésocentre de Calcul Intensif Aquitain (MCIA) and the IMN (Nantes) are acknowledged for computing facilities.

Table I. Comparison of experimental Fermi contact shifts (ppm, in bold) with the calculated ones obtained for various compounds with different functionals using either Eqs. (4a) or (4b) (see text). For the GGA+$U$, the effective potential ($U_{\text{eff}}$) is given in Ry and for the hybrid functionals (HyF1 of HyF2) mixing parameters are reported

| Compound | Site | GGA | GGA+$U$ ($U$eff) | | | HyF1($\alpha$) | | | HyF2($\alpha$) | | | **Exp** |
|---|---|---|---|---|---|---|---|---|---|---|---|---|
| | | | 0.10 | 0.30 | 0.50 | 0.10 | 0.35 | 0.50 | 0.10 | 0.35 | 0.50 | |
| LiMnPO$_4$ | Li | 131 | 120 | 105 | 93 | 125 | 90 | 112 | 107 | 56 | 24 | **68** |
| LiFePO$_4$ | Li | 118 | 85 | 68 | 58 | 92 | 70 | 65 | 59 | 9 | -2 | **-8** |
| LiCoPO$_4$ | Li | 73 | 39 | 31 | 27 | 36 | 26 | 24 | 8 | -21 | -20 | **-86** |
| LiNiPO$_4$ | Li | -63 | -53 | -40 | -30 | -62 | -22 | -15 | -60 | -44 | -33 | **-49** |
| | | | 0.10 | 0.30 | 0.50 | 0.10 | 0.15 | 0.30 | 0.10 | 0.15 | 0.30 | |
| Li$_6$CoO$_4$ | Li(1) | -176 | -160 | -103 | -110 | -152 | -142 | -90 | -242 | -265 | -307 | **-232** |
| | Li(2) | 1441 | 1300 | 1025 | 793 | 1198 | 1080 | 765 | 1092 | 941 | 605 | **885** |
| | | | 0.10 | 0.30 | 0.50 | 0.10 | 0.20 | 0.25 | 0.10 | 0.20 | 0.25 | |
| | Li(1) | 391 | 363 | 307 | 259 | 360 | 333 | 320 | 306 | 229 | 196 | **189** |
| Li$_3$Fe$_2$(PO$_4$)$_3$ | Li(2) | 222 | 174 | 165 | 137 | 199 | 179 | 170 | 149 | 86 | 58 | **89** |
| | Li(3) | 196 | 172 | 143 | 113 | 173 | 153 | 142 | 118 | 49 | 18 | **39** |
| | | | 0.10 | 0.22 | 0.30 | 0.10 | 0.20 | 0.25 | 0.10 | 0.20 | 0.25 | |
| | Li(1) | 135 | 138 | 128 | 122 | 129 | 131 | 129 | 97 | 104 | 99 | **103** |
| Li$_3$V$_2$(PO$_4$)$_3$ | Li(2) | 40 | 46 | 33 | 29 | 45 | 39 | 37 | 58 | 50 | 43 | **52** |
| | Li(3) | 65 | 36 | 30 | 27 | 33 | 17 | 14 | 37 | 24 | 16 | **17** |

FIGURE CAPTIONS

FIG. 1. (Color online) Environment of Li atoms in Li$M$PO$_4$. The LiO$_6$ octahedron shares edges with two $M$O$_6$ octahedra and corners with four other $M$O$_6$ polyhedra.

FIG. 2. (Color online) Comparison of experimental Fermi contact shifts of Li in olivine Li$M$PO$_4$ ($M$ = Mn, Fe, Co, Ni) with the calculated ones obtained with different approaches using Eq. (4a) (T = 310K). For the GGA+$U$, the effective potential ($U_{eff}$) is given in Ry and for the hybrid functionals (HyF1 of HyF2, see text) mixing parameters are reported: a = 0.35 (Mn, Co, and Ni); a = 0.5 (Fe).

FIG. 3. (Color online) Environment for the two Li sites of Li$_6$CoO$_4$ showing the local geometry: (a) Li(1); (b) Li(2).

FIG. 4. (Color online) Comparison of experimental Fermi contact shifts of Li(1) and Li(2) in Li$_6$CoO$_4$ with the calculated ones obtained with different approaches using Eq. (4a) (T = 320K). For the GGA+$U$, the effective potential ($U_{eff}$) is given in Ry and for the hybrid functionals (HyF1 of HyF2, see text) mixing parameters are reported.

FIG. 5. (Color online) Environment for the three Li sites of Li$_3$V$_2$(PO$_4$)$_3$ showing the local geometry: (a) Li(1); (b) Li(2); (c) Li(3).

FIG. 6. (Color online) Comparison of experimental Fermi contact shifts of Li(1), Li(2), and Li(3) in Li$_3M_2$(PO$_4$)$_3$ ($M$ = Fe (a); V (b)) with the calculated ones obtained with different approaches using Eqs. (4a) or (4b) (see text, T = 320K). For the GGA+$U$, the effective potential ($U_{eff}$) is given in Ry and for the hybrid functionals (HyF1 of HyF2, see text) mixing parameters are reported.

FIG. 7. (Color online) Average spin density around the three Li nuclei for $Li_3Fe_2(PO_4)_3$ (a) and $Li_3V_2(PO_4)_3$ (b) as a function of the integration sphere radius. Results are given for the GGA and GGA+$U$ ($U_{eff}$ = 3eV) calculations.

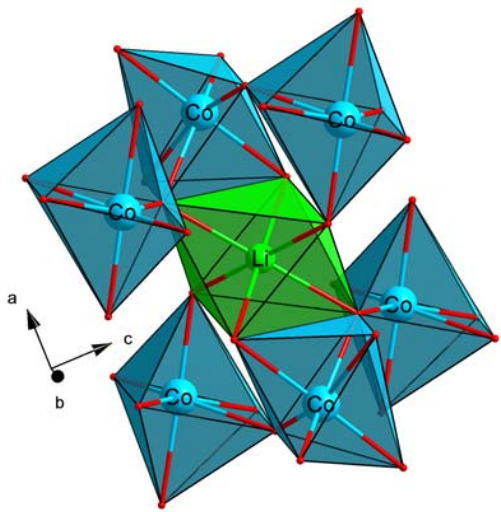

FIG. 1. (Color online) Environment of Li atoms in Li$M$PO$_4$. The LiO$_6$ octahedron shares edges with two $M$O$_6$ octahedra and corners with four other $M$O$_6$ polyhedra.

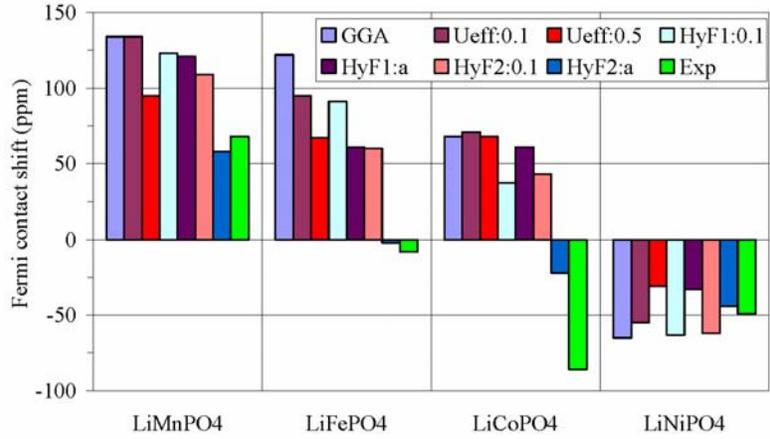

FIG. 2. (Color online) Comparison of experimental Fermi contact shifts of Li in olivine Li*M*PO$_4$ (*M* = Mn, Fe, Co, Ni) with the calculated ones obtained with different approaches using Eq. (4a) (T = 310K). For the GGA+*U*, the effective potential ($U_{\text{eff}}$) is given in Ry and for the hybrid functionals (HyF1 of HyF2, see text) mixing parameters are reported: a = 0.35 (Mn, Co, and Ni); a = 0.5 (Fe).

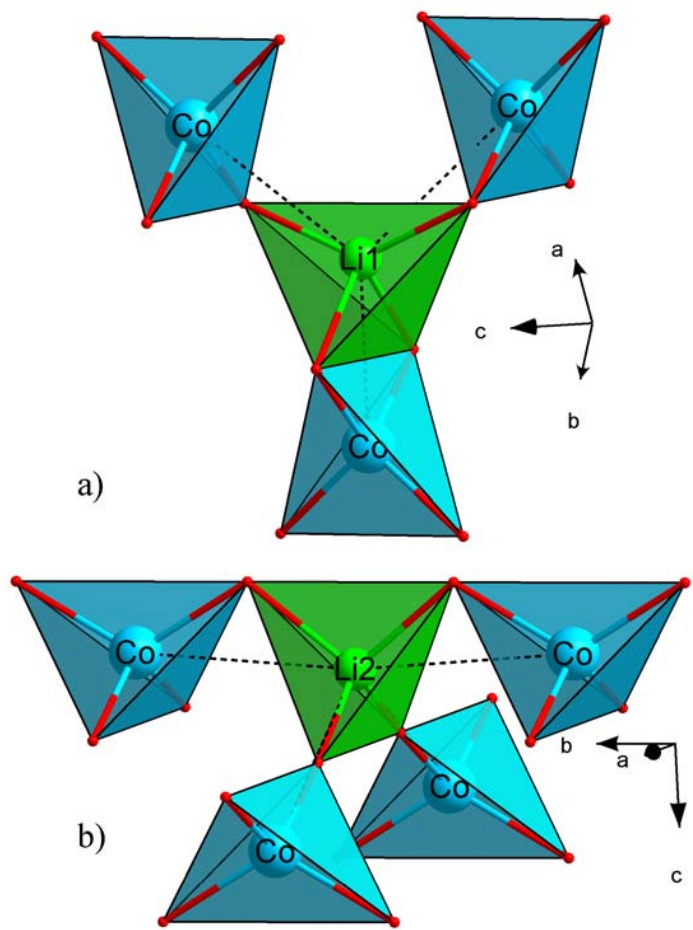

FIG. 3. (Color online) Environment for the two Li sites of $Li_6CoO_4$ showing the local geometry: (a) Li(1); (b) Li(2).

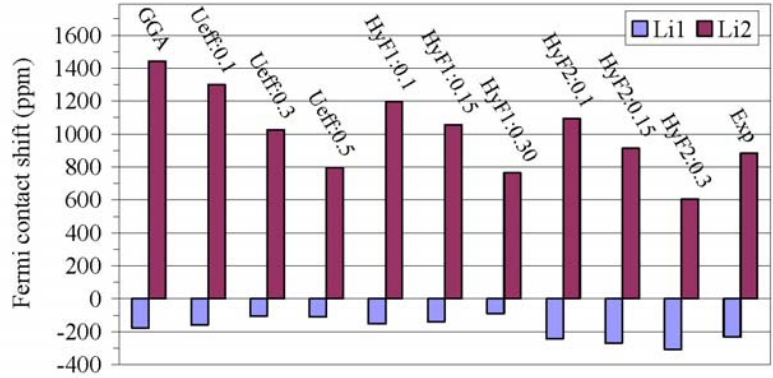

FIG. 4. (Color online) Comparison of experimental Fermi contact shifts of Li(1) and Li(2) in $Li_6CoO_4$ with the calculated ones obtained with different approaches using Eq. (4a) (T = 320K). For the GGA+$U$, the effective potential ($U_{eff}$) is given in Ry and for the hybrid functionals (HyF1 of HyF2, see text) mixing parameters are reported.

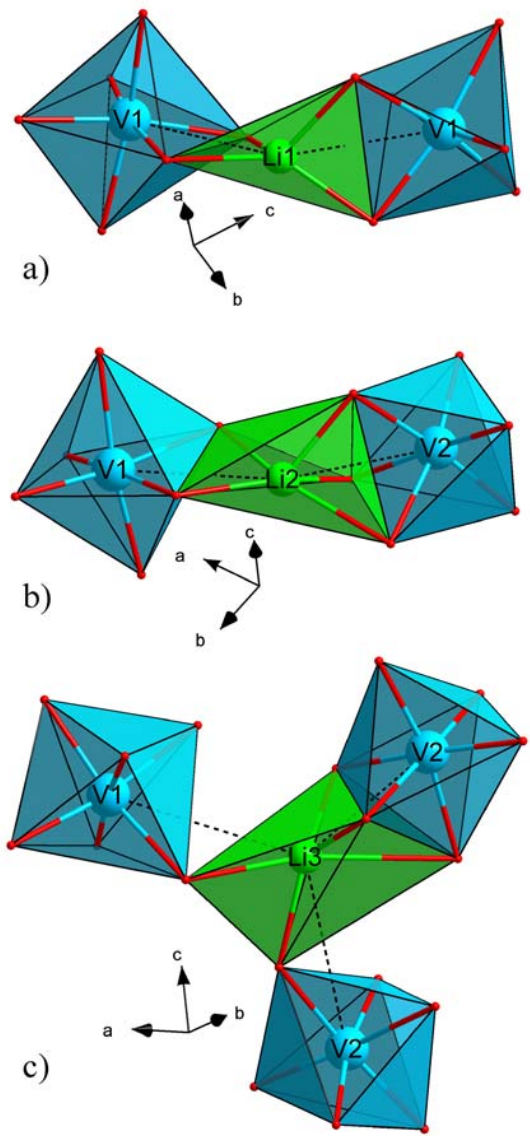

FIG. 5. (Color online) Environment for the three Li sites of $Li_3V_2(PO_4)_3$ showing the local geometry: (a) Li(1); (b) Li(2); (c) Li(3).

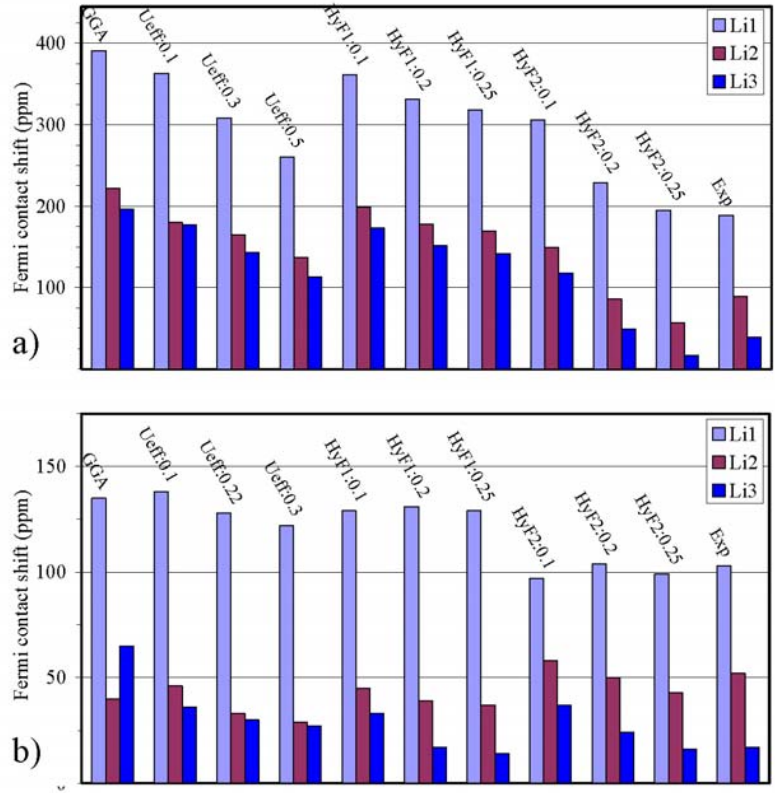

FIG. 6. (Color online) Comparison of experimental Fermi contact shifts of Li(1), Li(2), and Li(3) in $Li_3M_2(PO_4)_3$ ($M$ = Fe (a); V (b)) with the calculated ones obtained with different approaches using Eqs. (4a) or (4b) (see text, T = 320K). For the GGA+$U$, the effective potential ($U_{eff}$) is given in Ry and for the hybrid functionals (HyF1 of HyF2, see text) mixing parameters are reported.

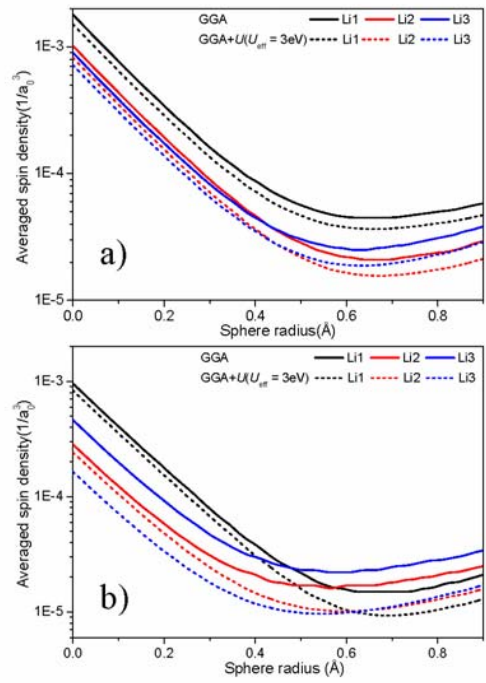

FIG. 7. (Color online) Average spin density around the three Li nuclei for $Li_3Fe_2(PO_4)_3$ (a) and $Li_3V_2(PO_4)_3$ (b) as a function of the integration sphere radius. Results are given for the GGA and GGA+$U$ ($U_{eff}$ = 3eV) calculations.